\documentclass[12pt]{article}

\title{\textbf{The Quantum Bit Commitment Theorem}}
\author{Jeffrey Bub\thanks{Email address: jbub@carnap.umd.edu} \\ 
\small \textit{Philosophy Department, University of Maryland, College 
Park, MD 20742, USA}}
\date{}

\normalsize
\begin{document}
\maketitle

\begin{abstract}
Unconditionally secure two-party bit commitment based \textit{solely} on 
the principles of quantum mechanics (without 
exploiting special relativistic signalling constraints, or principles of general 
relativity or 
thermodynamics) has been shown to be 
impossible, 
but the claim is repeatedly challenged. The quantum bit commitment theorem 
is reviewed 
here and the central conceptual 
point, that an `Einstein-Podolsky-Rosen' attack or cheating strategy 
can always be applied, is 
clarified. The question of whether following such a cheating strategy 
can ever be disadvantageous to the cheater is considered and answered 
in the negative. There is, indeed, no loophole in the theorem.
\end{abstract}  


\section{Introduction}

Over the past few years, the new fields of quantum information, 
quantum computation, and quantum cryptology have emerged as the locus 
of foundational research in quantum mechanics. In quantum 
cryptology, the main results have been a variety of provably secure 
protocols for key distribution, following the original Bennett and 
Brassard (BB84) protocol~\cite{BB84}, and an important `no go' theorem 
by Mayers~\cite{Mayers3,Mayers4,LC2}: the 
impossibility of unconditionally secure two-party bit commitment based 
\textit{solely} on the principles of quantum mechanics (without 
exploiting special relativistic signalling constraints, or principles of general 
relativity or 
thermodynamics). The 
quantum bit commitment theorem generalizes previous results restricted 
to one-way communication protocols by Mayers~\cite{Mayers2} 
and by Lo and 
Chau~\cite{LC1}, and applies to quantum, classical, and 
quantum-classical hybrid schemes (since classical information is 
essentially quantum information subject to certain constraints). The 
restriction to two-party schemes excludes schemes that involve a trusted 
third-party or trusted channel properties, and the restriction to 
schemes based solely on the principles of quantum mechanics excludes 
schemes that exploit special relativistic signalling 
constraints (see below), or schemes that might involve time 
machines or black holes.

In a key distribution protocol, the object is for two parties, Alice 
and Bob, who initially share no information, to exchange information 
via quantum and classical channels, so as to end up sharing a secret 
key (which they can then use for encryption), 
in such a way as to ensure that any attempt by an 
eavesdropper, Eve, to gain information about the secret key will be 
detected with non-zero probability.

The features of quantum mechanics that allow secure key distribution 
are, essentially, the quantum `no cloning' theorem (which makes it 
impossible for Eve to copy quantum communications between Alice and 
Bob for later analysis), and the fact that nonorthogonal quantum 
states cannot be distinguished without disturbing the states, so any 
information gain that depends on distinguishing such states 
must introduce some detectable disturbance. 

In a bit commitment protocol, one party, Alice, supplies an encoded bit 
to a second party, Bob. The information available in the encoding 
should be insufficient for Bob to ascertain the value of the bit, but 
sufficient, together with further information supplied by Alice at a 
subsequent stage when she is supposed to reveal the 
value of the bit, for Bob to be convinced that the protocol does not 
allow Alice to cheat by encoding the bit in a way that leaves her free 
to reveal either 0 or 1 at will. 

To illustrate the idea, suppose Alice 
claims the ability to predict advances or declines in the 
stock market on a daily basis. To substantiate her claim without 
revealing valuable information (perhaps to a potential 
employer, Bob) she suggests the following demonstration: She proposes 
to record her prediction, before the market opens, 
by writing a 0 (for `decline') or a 1 (for 
`advance') on a piece of paper, which she will lock in a safe. The 
safe will be handed to Bob, but Alice will keep the key. 
At the end of the day's trading, she will announce the bit she chose and prove  
that she in fact made the commitment at the earlier time by handing 
Bob the key. The question is whether there exists a quantum analogue 
of this procedure that is unconditionally secure: 
provably secure by the laws of physics against cheating by either Alice or 
Bob. Note that Bob can cheat if he can obtain \textit{some} 
information about Alice's commitment before she reveals it 
(which would give him an advantage in repetitions of the protocol with 
Alice). Alice can cheat if she can delay actually making a commitment 
until the final stage when she is required to reveal her commitment, 
or if she can change her commitment at the final stage with a very low 
probability of detection. 

The importance of quantum bit commitment as a cryptological primitive arises 
because of its relation to other cryptological protocols. 
Lo~\cite{Lo1} has argued that the 
impossibility of unconditionally secure quantum bit commitment implies 
the impossibility of secure quantum one-sided two-party computations, 
and hence the impossibility of secure quantum one-out-of-two 
oblivious transfer. It is easy to see that a remote coin 
tossing procedure, in which neither party can cheat, would be possible 
if secure bit 
commitment were possible, which would allow unconditionally secure remote 
gambling (gambling over the internet, for example). But note that a procedure 
for remote fair games has been proposed by Goldenberg, Vaidman, and 
Wiesner~\cite{GVW}, 
so this is a 
weaker protocol than bit commitment.

Bennett and Brassard originally proposed a quantum bit commitment protocol 
in~\cite{BB84}. The basic idea was to associate the 0 and 1 
commitments with two statistically equivalent quantum mechanical mixtures 
(represented by the same density operator). As they showed in the 
same paper, Alice can cheat by adopting an `Einstein-Podolsky-Rosen' 
(EPR) 
attack or cheating strategy: she prepares entangled pairs of 
particles, keeps one of each pair (the ancilla) and sends the second 
particle (the channel particle) to Bob. In this way she can fake 
sending one of two equivalent mixtures to Bob 
and reveal either bit at will at the opening stage by 
effectively creating
the desired mixture via appropriate measurements on her ancillas. Bob 
cannot detect this cheating strategy. 

In a later paper~\cite{BCJL}, 
Brassard, Cr\'{e}peau, Josza, and Langlois proposed a bit commitment 
protocol that they 
claimed to be unconditionally secure. The BCJL scheme was first shown 
to be insecure by Mayers~\cite{Mayers1,Mayers2}. Subsequently, 
Mayers~\cite{Mayers3,Mayers4} and Lo and 
Chau~\cite{LC1} independently showed that a large class of quantum 
bit commitment schemes are insecure. Lo and Chau presented their 
result in
\cite{LC1} as applicable only to all proposed quantum bit commitment 
schemes, including the BCJL scheme (for which they  relied on Mayers' 
extended analysis in \cite{Mayers2}). But as Mayers showed in 
\cite{Mayers3,Mayers4}, the insight of 
Bennett and Brassard
in \cite{BB84} can be extended to a proof that a generalized version of 
the EPR cheating strategy can always be applied, if the 
Hilbert space is enlarged in a suitable way by introducing additional 
ancilla particles. Following Mayers, a similar result is proved in Lo 
and Chau~\cite{LC2}, where the operative assumption is that both 
Alice and Bob have available quantum computers of unlimited power and 
are capable of storing quantum signals indefinitely. 

Mayers' analysis in \cite{Mayers4} explicitly models 
the exchange of quantum and classical information in two-way quantum bit 
commitment protocols via a `direct' approach. For an interesting 
`indirect' or `reduction' approach, see \cite{BCMS1,BCM,BCMS2}.
Classical information can be understood as a type of quantum information 
with additional constraints. The distinction between classical and quantum 
information was always explicit 
in the analysis of proposed quantum bit commitment protocols
According to Mayers (personal 
communication), this explains why
researchers failed to see the general impossibility of quantum bit 
commitment, even
after the basic mathematical result, which is valid in a purely quantum 
world, was known. 

The negative 
results of Mayers and Lo and Chau came as a surprise and 
were received with dismay by 
the quantum cryptology community. The proof of the basic theorem, 
which exploits the biorthogonal decomposition 
theorem, is remarkably simple, but the impossibility of secure bit 
commitment based solely on the principles of quantum (or classical) mechanics 
has profound consequences. Indeed, it would not 
be an exaggeration to say that the significance of the quantum bit commitment 
theorem is comparable to Bell's locality theorem \cite{Bell} for 
quantum mechanics. Brassard and Fuchs have 
speculated (private communication and \cite{Fuchs}) 
that quantum mechanics can be derived from two 
postulates about quantum information: the possibility of secure 
key distribution and the impossibility of secure bit commitment. 
That is, in a quantum world the communication of information is 
characterized precisely in this way in terms of a limited sort of privacy. 

Perhaps because of the simplicity of the proof and the 
universality of the claim, the quantum bit commitment theorem is 
continually challenged in the literature (see, for example, 
\cite{Cheung,Mitra,Yuen}), 
on the basis that the proof does not cover all possible procedures 
that might be exploited to implement quantum bit commitment. 
There seems to be a 
general feeling that the theorem is `too good to be true' and that 
there must be a loophole. 

In fact, there is no loophole. While Kent~\cite{Kent1,Kent2} has shown 
how to implement a secure classical 
bit commitment protocol by exploiting relativistic signalling 
constraints in a timed sequence of communications between verifiably 
separated sites for both Alice and Bob, 
and Hardy and Kent~\cite{HardyKent} and Aharonov, Ta-Shma, Vazirani, 
and Yao~\cite{Aharonov} have investigated the security of 
`cheat-sensitive' or `weak' versions of quantum bit commitment, 
these results are not in conflict with the quantum bit commitment 
theorem. 
In a bit commitment protocol as usually construed, there is a time interval 
of arbitrary 
length, where no information is exchanged, 
between the end of the commitment stage of the protocol and 
the opening or unveiling stage, when Alice reveals the value of the 
bit. Kent's ingenious scheme effectively involves a third stage between the 
commitment stage and the unveiling stage, in which information is 
exchanged between Bob's sites and Alice's sites at regular intervals 
until one of Alice's sites 
chooses to unveil the originally committed bit. At this moment of 
unveiling the protocol is not yet complete, because a further sequence of 
unveilings is required between Alice's sites and corresponding sites 
of Bob before Bob has all the information required to verify the 
commitment at a single site. If a bit commitment protocol 
is understood to 
require an arbitrary amount of `free' time between the end of the 
commitment stage and the opening stage (in which no step is to be 
executed in the protocol), then the quantum bit commitment theorem 
covers protocols that exploit special relativistic signalling 
constraints.  (I am indebted to Dominic Mayers for clarifying this 
point.)
The aim of the 
following discussion will be to clarify the underlying 
logic of the proof, and especially the crucial significance of the 
assumption
that both parties can be assumed to have access to 
quantum computers, so that a (generalized) EPR cheating 
strategy is always possible.

In Section 2, I review the structure of the proof and show how any 
step in a bit commitment protocol that requires Alice or Bob to make a 
determinate choice (whether to perform one of a number of alternative 
measurements, or whether to implement one of a number of alternative 
unitary transformations) can always be replaced by an EPR cheating 
strategy in the generalized sense, assuming that Alice and Bob are 
both equipped with quantum computers. That is, a classical 
disjunction over determinate possibilities---this operation 
\textit{or} that operation---can always be replaced by a quantum 
entanglement and a subsequent measurement (perhaps at a more 
convenient time for the cheater) in which one of the possibilities 
becomes determinate. Essentially, the classical disjunction is 
replaced by a quantum disjunction. This cheating strategy cannot be 
detected. Similarly, a measurement can be `held at the quantum level' 
without detection: instead of performing the measurement and obtaining 
a determinate outcome as one of a number of possible outcomes, a 
suitable unitary transformation can be performed on an enlarged 
Hilbert space, in which the system is entangled with a `pointer' 
ancilla in an appropriate way, and the procedure of obtaining a 
determinate outcome (which involves decoherence, or the `collapse' of 
the quantum state onto an eigenstate of the observable measured) can 
be delayed. The possibility of keeping the series of transactions 
between Alice and Bob at the quantum level by enlarging the 
Hilbert space, until the final exchange of classical information when 
Alice reveals her commitment, is the crucial insight that underlies 
Mayers' general proof. In John Smolin's whimsical terminology, 
this is the doctrine of the Church of the Larger 
Hilbert Space: the belief that a fully quantum treatment can always be 
obtained by extending the Hilbert space.

If it can be assumed that a measurement has in fact been performed 
and a determinate outcome obtained, then secure bit commitment is 
possible. This is tantamount to assuming that an EPR cheating strategy is 
blocked. Since there is no way to distinguish whether the protocol has 
been followed or replaced by an EPR cheating strategy, it would seem 
that there is no way to ensure that a measurement has in fact been 
performed and a determinate outcome recorded. 

But how do we know that 
there is no bit commitment protocol of 
the following sort: Suppose, at some stage of the protocol, Bob (say) 
is required to perform one of two alternative measurements, X or Y, chosen 
at random. If Bob 
actually chooses one of X or Y, and actually performs the measurement 
and obtains a determinate outcome, then the protocol is secure 
against cheating by both parties. If Bob implements an EPR strategy 
and keeps the choice and the 
measurement at the quantum level, 
then it turns out that Alice has a greater probability 
of cheating successfully than Bob. If there were such a protocol, 
then even though Bob could implement an EPR strategy without 
detection, he would effectively be forced to make the choice and carry out the 
measurement, since he would not 
choose to put himself in a weaker position relative to Alice over the 
long run in a series of bit commitment transactions. 
In Section 3, I show that the possibility of such a protocol 
is blocked by the theorem itself. That is, adopting an EPR cheating strategy 
is never disadvantageous to the cheater.

\section{The Bit Commitment Theorem}

Any bit commitment scheme will involve a series of transactions 
between Alice and Bob, where a certain number, $n$, of quantum 
systems---the `channel particles'---are passed between them and
subjected to various operations (unitary transformations, 
measurements), possibly chosen randomly. I show now how these 
operations can always be replaced, without detection, 
by entangling a channel particle with one or more ancilla particles 
that function as `pointer' particles for measurements or `dice' 
particles for random choices. This is the (generalized) EPR cheating strategy.

Suppose, at a certain stage of a bit commitment protocol, 
that Bob is required to make a random choice between 
measuring one of two observables, $X$ or $Y$, on each channel particle he 
receives from Alice. For simplicity, assume that $X$ and $Y$ each have two 
eigenvalues, $x_{1}$, $x_{2}$ and $y_{1}$, $y_{2}$.
After recording the outcome of the measurement, 
Bob is required to return the channel particle to Alice. When Alice 
receives the $i$'th channel particle she sends Bob the next channel particle 
in the sequence. We may suppose that the measurement outcomes that Bob 
records form part of the information that enables him to confirm 
Alice's commitment, once she discloses it (together with further 
information), so he is not required to report his measurement outcomes 
to Alice until the final stage of the protocol when she reveals her commitment.

Instead of following the protocol, Bob can construct a device that 
entangles the input state $|\psi\rangle_{C}$ of a 
channel particle with the initial 
states, $|d_{0}\rangle_{B}$ and $|p_{0}\rangle_{B}$, of two ancilla 
particles that he introduces, 
the first of which functions as a `quantum die' for the 
random choice and the 
second as a `quantum pointer' for the measurement. It is assumed that 
Bob's ability to construct such a device---a special purpose quantum 
computer---is restricted only by the laws of quantum mechanics. 
The entanglement  
is implemented by a unitary transformation in the following way:\footnote{Note 
that there is 
no loss of generality in assuming that the channel particle is in a 
pure state. If the channel particle is entangled with Alice's 
ancillas, the device implements the entanglement via the 
transformation $I\otimes \cdots$, 
where $I$ is the identity operator in the Hilbert space of Alice's 
ancillas.} 
Define two unitary transformations, $U_{X}$ and $U_{Y}$, that 
implement the $X$ and $Y$ measurements `at the quantum level' on the 
tensor product of the Hilbert space of the channel particle, 
$\mathcal{H}_{C}$, and the Hilbert space of 
Bob's pointer ancilla, $\mathcal{H}_{B(P)}$: 
\begin{eqnarray}
    |x_{1}\rangle_{C}|p_{0}\rangle_{B} \stackrel{U_{X}}{\longrightarrow} 
    |x_{1}\rangle_{C}|p_{1}\rangle_{B}
    \nonumber \\
    |x_{2}\rangle_{C}|p_{0}\rangle_{B} \stackrel{U_{X}}{\longrightarrow} 
    |x_{2}\rangle_{C}|p_{2}\rangle_{B}
\end{eqnarray}
and
\begin{eqnarray}
    |y_{1}\rangle_{C}|p_{0}\rangle_{B} \stackrel{U_{Y}}{\longrightarrow} 
    |y_{1}\rangle_{C}|p_{1}\rangle_{B}
    \nonumber \\
    |y_{2}\rangle_{C}|p_{0}\rangle_{B} \stackrel{U_{Y}}{\longrightarrow} 
    |y_{2}\rangle_{C}|p_{2}\rangle_{B}
\end{eqnarray}
so that 
\begin{equation}
    |\psi\rangle_{C}|p_{0}\rangle_{B} 
    \stackrel{U_{X}}{\longrightarrow}
    \langle x_{1}|\psi\rangle|x_{1}\rangle_{C} |p_{1}\rangle_{B}
    + \langle x_{2}|\psi\rangle|x_{2}\rangle_{C} |p_{2}\rangle_{B}
\end{equation}
and
\begin{equation}
    |\psi\rangle_{C}|p_{0}\rangle_{B} 
    \stackrel{U_{Y}}{\longrightarrow}
    \langle y_{1}|\psi\rangle|y_{1}\rangle_{C} |p_{1}\rangle_{B}
    + \langle y_{2}|\psi\rangle|y_{2}\rangle_{C} |p_{2}\rangle_{B}
\end{equation}

The random choice is defined similarly by a unitary transformation 
$V$ on 
the tensor product of the Hilbert space of Bob's die ancilla, 
$\mathcal{H}_{B(D)}$, and the Hilbert space 
$\mathcal{H}_{C}\otimes\mathcal{H}_{B(P)}$. Suppose $|d_{X}\rangle$ 
and $|d_{Y}\rangle$ are two orthogonal states in $\mathcal{H}_{B(D)}$ 
and that $|d_{0}\rangle = \frac{1}{\sqrt{2}}|d_{X}\rangle + 
\frac{1}{\sqrt{2}}|d_{Y}\rangle$. Then (suppressing 
the obvious subscripts) $V$ is defined by:
\begin{eqnarray}
    |d_{X}\rangle\otimes |\psi\rangle|p_{0}\rangle &
    \stackrel{V}{\longrightarrow} & |d_{X}\rangle \otimes 
    U_{X}|\psi\rangle|p_{0}\rangle \nonumber \\
    |d_{Y}\rangle\otimes |\psi\rangle|p_{0}\rangle &
    \stackrel{V}{\longrightarrow} & |d_{Y}\rangle \otimes 
    U_{Y}|\psi\rangle|p_{0}\rangle
\end{eqnarray}
so that
\begin{eqnarray}
    \lefteqn{|d_{0}\rangle \otimes|\psi\rangle|p_{0}\rangle
    \stackrel{V}{\longrightarrow}} \nonumber \\
    & & \frac{1}{\sqrt{2}}|d_{X}\rangle \otimes 
    U_{X}|\psi\rangle|p_{0}\rangle 
      + \frac{1}{\sqrt{2}}|d_{Y}\rangle \otimes
     U_{Y}|\psi\rangle|p_{0}\rangle
     \label{eq:V}
\end{eqnarray}
where the tensor product symbol has been introduced selectively to indicate 
that $U_{x}$ and $U_{y}$ are defined on 
$\mathcal{H}_{C}\otimes\mathcal{H}_{B(P)}$.

If Bob were to actually 
choose the observable $X$ or $Y$ randomly, and actually perform the 
measurement and obtain a particular eigenvalue, Alice's density operator for 
the channel particle would be:
\begin{eqnarray}
    \lefteqn{\frac{1}{2}(|\langle x_{1}|\psi\rangle|^{2}|x_{1}\rangle\langle x_{1}|
    + |\langle x_{2}|\psi\rangle|^{2}|x_{2}\rangle\langle x_{2}|)} 
    \nonumber \\
    & & \frac{1}{2}(|\langle y_{1}|\psi\rangle|^{2}|y_{1}\rangle\langle y_{1}|
    + |\langle y_{2}|\psi\rangle|^{2}|y_{2}\rangle\langle y_{2}|)
    \label{eq:density1}
\end{eqnarray}
assuming that Alice does not know what observable Bob chose to 
measure, nor what outcome he obtained.
But this is precisely the same density operator generated by tracing 
over Bob's ancilla particles for the state produced in (\ref{eq:V}). 
In other words, the density operator for 
the channel particle is the same for Alice, whether Bob 
randomly chooses which observable to measure and actually performs 
the measurement, or whether he implements an EPR cheating strategy 
with his two ancillas that produces the transition (\ref{eq:V}) on the 
enlarged Hilbert space. 

If Bob is required to eventually report what measurement he performed 
and what outcome he obtained, he can at that stage measure the die 
ancilla for the eigenstate $|d_{X}\rangle$ or $|d_{Y}\rangle$, and then 
measure the pointer ancilla for the eigenstate $|p_{1}\rangle$ or 
$|p_{2}\rangle$. In effect, if we consider the ensemble of possible 
outcomes for the two measurements, Bob will have converted the 
`improper' mixture generated by tracing over his ancillas to a `proper' 
mixture. But the difference between a proper and improper mixture 
is undetectable by Alice since she has no access to Bob's ancillas, 
and it is only by measuring the composite system consisting of the 
channel particle together with Bob's ancillas that Alice could 
ascertain that the channel particle is entangled with the ancillas. 

In fact, if it were possible to distinguish between a proper and 
improper mixture, it would be possible to signal superluminally: 
Alice could know instantaneously whether or not Bob performed a 
measurement on his ancillas by monitoring the channel particles in her 
possession. Note that it makes no difference whether Bob or Alice 
measures first, since the measurements are of observables in 
different Hilbert spaces, which therefore commute. 

Clearly, a similar argument applies if Bob is required to choose between 
alternative unitary operations at some stage of a bit commitment 
protocol. Perhaps less obviously, an EPR cheating strategy is also possible if 
Bob is required to perform a measurement or choose between alternative operations
on channel particle $i+1$, 
conditional on the outcome 
of a prior measurement on channel particle $i$, or conditional on a 
prior choice of some operation from among a set of alternative 
operations. Of course, if Bob is in possession of all the channel particles 
at the same time, he can perform an entanglement with ancillas on the entire 
sequence, considered as a single composite system. But even if Bob 
only has access to one channel particle at a time (which he is 
required to return to Alice after performing a measurement or other operation 
before 
she sends him the next channel particle), he can always entangle 
channel particle $i+1$ with the 
ancillas he used to entangle channel 
particle $i$. 

For example, suppose Bob is presented with two channel particles in 
sequence. He is supposed to decide randomly whether to measure $X$ or 
$Y$ on the first particle, perform the measurement, and return the 
particle to Alice. After Alice receives the first particle, she sends 
Bob the second particle. If Bob measured $X$ on the first particle and 
obtained the outcome $x_{1}$, he 
is supposed to measure $X$ on the second particle; if he obtained 
the outcome $x_{2}$, he is supposed to measure $Y$ on the second 
particle. If he measured $Y$ on the first particle and obtained the 
outcome $y_{1}$, he is supposed to apply the unitary transformation 
$U_{1}$ to the second particle; if he obtained the outcome $y_{2}$, he 
is supposed to apply the unitary transformation $U_{2}$. After 
performing the required operation, he is supposed to return the 
second particle to Alice. 

It would seem at first sight that Bob has to 
actually perform a measurement on the first channel particle and obtain a 
particular outcome before he can apply the protocol to the second 
particle, given that he only has access to one channel particle at a time, 
so an EPR cheating strategy is excluded. But this is not so. Bob's strategy is 
the following: He applies the EPR strategy discussed above for two 
alternative measurements to the first 
channel particle. 
For the second channel particle, he applies the following unitary 
transformation on the tensor product of the Hilbert spaces of his 
ancillas and the channel particle, where the state of the second channel 
particle is denoted by $|\phi\rangle$, 
and the state of the pointer ancilla for the second channel particle 
is denoted by $|q_{0}\rangle$ (a second die particle is not required):

\begin{eqnarray}
    |d_{X}\rangle|p_{1}\rangle |\phi\rangle |q_{0}\rangle 
    \stackrel{U_{C}}{\longrightarrow}
    |d_{X}\rangle |p_{1}\rangle \otimes U_{X} |\phi\rangle|q_{0}\rangle 
    \nonumber \\
    |d_{X}\rangle|p_{2}\rangle |\phi\rangle |q_{0}\rangle 
    \stackrel{U_{C}}{\longrightarrow}
    |d_{X}\rangle |p_{2}\rangle \otimes U_{Y} |\phi\rangle|q_{0}\rangle
    \nonumber \\
    |d_{Y}\rangle |p_{1}\rangle|\phi\rangle |q_{0}\rangle
    \stackrel{U_{C}}{\longrightarrow} 
    |d_{Y}\rangle|p_{1}\rangle \otimes U_{1}|\phi\rangle \otimes |q_{0}\rangle
    \nonumber \\
    |d_{Y}\rangle |p_{2}\rangle|\phi\rangle |q_{0}\rangle
    \stackrel{U_{C}}{\longrightarrow} 
    |d_{Y}\rangle|p_{2}\rangle \otimes U_{2}|\phi\rangle \otimes |q_{0}\rangle
\end{eqnarray}

Since an EPR cheating strategy can always be applied without detection, the 
proof of the bit commitment theorem assumes that at the end of the 
commitment stage the composite system consisting of Alice's ancillas, 
the $n$ channel particles, and Bob's
ancillas will be represented by some composite entangled 
state $|0\rangle$ or $|1\rangle$, depending on Alice's 
commitment, on a Hilbert space 
$\mathcal{H}_{A}\otimes \mathcal{H}_{B}$, where 
$\mathcal{H}_{A}$ is the Hilbert space of the particles in Alice's 
possession at that stage (Alice's ancillas and the channel particles 
retained by Alice, if any), and 
$\mathcal{H}_{B}$ is the Hilbert space of the particles in Bob's 
possession at that stage (Bob's ancillas and the channel particles 
retained by Bob, if any). 

Now, the density operators $W_{B}(0)$ and 
$W_{B}(1)$ characterizing the information available to 
Bob for the two alternative commitments 
are obtained by tracing the states $|0\rangle$ and $|1\rangle$ 
over $\mathcal{H}_{A}$. If these density operators are the same, 
then Bob will be unable to distinguish the 
0-commitment from the 1-commitment without further information from 
Alice. In this case, the protocol is said to be `concealing.' What the proof 
establishes, by an application of the biorthogonal decomposition 
theorem, is that if $W_{B}(0) = W_{B}(1)$ then there exists a unitary 
transformation in $\mathcal{H}_{A}$ that will transform $|0\rangle$ 
to $|1\rangle$. That is, if the protocol is `concealing' then 
it cannot be `binding' on Alice: she can always make the 0-commitment 
and follow the protocol (with appropriate applications of an EPR  
strategy) to establish the state $|0\rangle$. At the 
final stage when she is required to reveal her commitment, she can 
change her commitment if she chooses, depending on circumstances, by applying a 
suitable unitary transformation in her own Hilbert space to transform 
$|0\rangle$ to $|1\rangle$ without Bob being able to 
detect this move. So 
either Bob can cheat by obtaining some information about Alice's 
choice before she reveals her commitment, or Alice can cheat. 

The essentials of the proof can be sketched as follows: In the 
biorthogonal (Schmidt) decomposition, the states $|0\rangle$ and 
$|1\rangle$ can be expressed as:
\begin{eqnarray}
    |0\rangle & = & \sum_{i}\sqrt{c_{i}}|a_{i}\rangle|b_{i}\rangle
    \nonumber \\
    |1\rangle & = & \sum_{j}\sqrt{c'_{j}}|a'_{j}\rangle|b'_{j}\rangle
\end{eqnarray}
where $\{|a_{i}\rangle\}, \{|a'_{j}\rangle\}$ are two 
orthonormal sets of states in $\mathcal{H}_{A}$, and 
$\{|b_{i}\rangle\}, \{|b'_{j}\rangle\}$ are two orthonormal 
sets in $\mathcal{H}_{B}$.

The density operators $W_{B}(0)$ and $W_{B}(1)$ are defined by:
\begin{eqnarray}
    W_{B}(0) = Tr_{A}|0\rangle\langle 0| & = & 
    \sum_{i}c_{i}|b_{i}\rangle\langle b_{i}|
    \nonumber \\
     W_{B}(1) = Tr_{A}|1\rangle\langle 1| & = & 
    \sum_{j}c'_{j}|b'_{j}\rangle\langle b'_{j}|
\end{eqnarray}

Bob can't cheat if and only if $W_{B}(0) = W_{B}(1)$. Now, by the spectral 
theorem, the decompositions:
\begin{eqnarray*}
    W_{B}(0) & = & \sum_{i}c_{i}|b_{i}\rangle\langle b_{i}|
    \nonumber \\
    W_{B}(1) & = & \sum_{j}c'_{j}|b'_{j}\rangle\langle b'_{j}|
\end{eqnarray*}
are unique. For the nondegenerate case, where the $c_{i}$ are all 
distinct and the $c'_{j}$ are all distinct, the condition $W_{B}(0) = 
W_{B}(1)$ implies that for all $k$:
\begin{eqnarray}
    c_{k} & = & c'_{k} 
    \nonumber \\
    |b_{k}\rangle & = & |b'_{k}\rangle
\end{eqnarray}
and so
\begin{eqnarray}
    |0\rangle & = & \sum_{k}\sqrt{c_{k}}|a_{k}\rangle|b_{k}\rangle
    \nonumber \\
    |1\rangle & = & \sum_{k}\sqrt{c_{k}}|a'_{k}\rangle|b_{k}\rangle
\end{eqnarray}
It follows that there exists a unitary transformation $U\in \mathcal{H}_{A}$ 
such that
\begin{equation}
    \{|a_{k}\rangle\} \stackrel{U}{\longrightarrow} \{|a'_{k}\rangle\}
\end{equation}
and hence
\begin{equation}
    |0\rangle \stackrel{U}{\longrightarrow} |1\rangle
\end{equation}
    
The degenerate case can be handled in a similar way. Suppose that 
$c_{1} = c_{2} = c'_{1} = c'_{2} = c$. Then $|b_{1}\rangle, 
|b_{2}\rangle$ and $|b'_{1}\rangle, 
|b'_{2}\rangle$ span the same subspace $\mathcal{H}$ in 
$\mathcal{H}_{B}$, and hence (assuming the coefficients are distinct 
for $k>2$:
\begin{eqnarray}
    |0\rangle & = &  
    \sqrt{c}(|a_{1}\rangle|b_{1}\rangle + |a_{2}\rangle|b_{2}\rangle)
    + \sum_{k>2}\sqrt{c_{k}}|a_{k}\rangle|b_{k}\rangle
    \nonumber \\
    |1\rangle & = &  
    \sqrt{c}(|a'_{1}\rangle|b'_{1}\rangle + |a'_{2}\rangle|b'_{2}\rangle)
    + \sum_{k>2}\sqrt{c_{k}}|a'_{k}\rangle|b_{k}\rangle
    \nonumber \\
    & = & 
    \sqrt{c}(|a''_{1}\rangle|b_{1}\rangle + |a''_{2}\rangle|b_{2}\rangle
    + \sum_{k>2}\sqrt{c_{k}}|a'_{k}\rangle|b_{k}\rangle
\end{eqnarray}
where $|a''_{1}\rangle, |a''_{2}\rangle$ are orthonormal states 
spanning $\mathcal{H}$. Since $\{|a''_{1}\rangle, 
|a''_{2}\rangle, |a_{3}\rangle, \ldots \}$ is an orthonormal set in 
$\mathcal{H}_{A}$, there exists a unitary transformation in 
$\mathcal{H}_{A}$ that transforms $\{|a_{k}\rangle\}$ to 
$\{|a''_{1}\rangle, |a''_{2}\rangle, |a'_{3}\rangle, \dots\}$, 
and hence $|0\rangle$ to $|1\rangle$ 

The extension of the theorem to the nonideal case, where $W_{B}(0) 
\approx W_{B}(1)$, so that there is a small probability of Bob 
distinguishing the 
alternative commitments, shows that Alice has a correspondingly large 
probability of cheating successfully: there exists a $U$ that will 
transform $W_{B}(0)$ sufficiently close to $W_{B}(1)$ so that Bob has a very 
small probability of making the distinction.  

The heart of the mathematical proof is the biorthogonal 
decomposition theorem. But the essential conceptual insight is the possibility 
of enlarging the Hilbert space and implementing an EPR 
strategy without detection. This raises the 
following question, considered in the next section: Suppose Bob cannot 
cheat because $W_{B}(0) = W_{B}(1)$, so by the theorem there exists a 
unitary transformation $U$ in $\mathcal{H}_{A}$ that will transform
$|0\rangle$ to $|1\rangle$. Could there be a protocol 
in which Alice also cannot cheat because, although there exists a 
suitable unitary transformation $U$, 
she cannot know what unitary 
transformation to apply? In the next section we shall see that this 
is indeed the case, but only if $U$ 
depends on Bob's operations, which are unknown to Alice. But then Bob 
would have to actually make a determinate choice or obtain a 
determinate outcome in a measurement, and he could always avoid doing 
so without detection by applying an EPR strategy. The remaining 
question would seem to be whether he might choose to avoid an EPR 
strategy in a certain situation because it would be disadvantageous 
to him. How do we know that following an EPR strategy is never 
disadvantageous?

\section{A Possible Loophole?}

The question at issue in this section is whether applying an EPR cheating 
strategy can ever be disadvantageous to the cheater. Note that the 
standard approach in cryptology 
is to consider the possibility of cheating against an honest opponent. 
Here we are considering the question of whether a quantum bit 
commitment protocol exists with the feature that one of the parties 
would forego a certain cheating strategy, because the opposing party 
would be able to cheat by taking advantage of such a move. So, 
strictly speaking, this would not be considered a loophole in the 
quantum bit commitment theorem, even if we could identify such a 
protocol. Nevertheless, this `game-theoretic' extension 
of the usual notion is certainly relevant to the issue of 
security.

To focus the question, it will be worthwhile to consider a 
particular protocol based on the Aharonov-Bergmann-Lebowitz notion of 
pre- and post-selected quantum states~\cite{ABL}. If (i) Alice 
prepares a system in a 
certain state $|\mbox{pre}\rangle$ 
at time $t_{1}$, (ii) Bob measures some observable $Q$ on the system 
at time $t_{2}$, and (iii) Alice measures an observable of which 
$|\mbox{post}\rangle$ is an eigenstate at time $t_{3}$, 
and post-selects for $|\mbox{post}\rangle$, then Alice can assign 
probabilities to the outcomes of Bob's $Q$-measurement at $t_{2}$, 
conditional on the states $|\mbox{pre}\rangle$ and $|\mbox{post}\rangle$ at 
times 
$t_{1}$ and $t_{3}$, respectively, as follows:
\begin{equation}
    \mbox{prob}(q_{k}) =
    \frac{|\langle \mbox{pre}|P_{k}| \mbox{post}\rangle|^{2}}
    {\sum_{i} |\langle \mbox{pre} |P_{i}|\mbox{post}\rangle|^{2}}
\end{equation}
where $P_{i}$ is the projection operator onto the $i$'th eigenspace 
of $Q$. Notice that the $ABL$-rule is 
time-symmetric, in the sense that the states $|\mbox{pre}\rangle$ and 
$|\mbox{post}\rangle$ can be interchanged, so these states are 
sometimes referred to as time-symmetric states.

If $Q$ is unknown to Alice, she can use this `ABL-rule' to assign 
probabilities to the outcomes of various hypothetical 
$Q$-measurements. The interesting peculiarity of the ABL-rule, by 
contrast with the usual Born rule for pre-selected states, is that it 
is possible---for an appropriate choice of observables $Q$, $Q'$, 
\ldots, and states $|\mbox{pre}\rangle$ and $|\mbox{post}\rangle$---to 
assign unit probability to the outcomes of a set of mutually 
\textit{noncommuting} observables. That is, Alice can be in a 
position to assert a conjunction of conditional statements of the 
form: `If Bob measured $Q$, then the outcome must have been $q_{i}$, 
with certainty, and if Bob measured $Q'$, then the outcome must have been 
$q'_{j}$, with certainty, \ldots,' where $Q, Q', \ldots$ are mutually 
noncommuting observables.   

A case of this sort has been discussed by 
Vaidman, Aharonov, and Albert~\cite{VAA}, where the outcome of a 
measurement of any of the three spin components $\sigma_{x}$, 
$\sigma_{y}$, $\sigma_{z}$ of a spin-$\frac{1}{2}$ particle can be 
inferred from an appropriate pre- and post-selection. Alice prepares 
a pair of particles, $A$ and $C$, in the Bell state:
\begin{equation}
    |\mbox{pre}\rangle = 
    \frac{1}{\sqrt{2}}(|\uparrow_{z}\rangle_{A}|\uparrow_{z}\rangle_{C} + 
    |\downarrow_{z}\rangle_{A}|\downarrow_{z}\rangle_{C}
    \label{eq:Bell}
\end{equation}
where $|\uparrow_{z}\rangle$ and $|\downarrow_{z}\rangle$  
denote the $\sigma_{z}$-eigenstates. Alice sends the channel particle
$C$ to Bob 
and keeps the ancilla $A$. Bob measures either 
$\sigma_{x}$, or $\sigma_{y}$, or $\sigma_{z}$ on the channel 
particle and returns the channel particle to Alice. Alice then 
measures an observable $R$ on the pair of particles, where $R$ has 
the eigenstates:
\begin{eqnarray}
    |r_{1}\rangle & = & 
    \frac{1}{\sqrt{2}}|\uparrow_{z}\rangle|\uparrow_{z}\rangle + 
    \frac{1}{2}(|\uparrow_{z}\rangle|\downarrow_{z}\rangle 
    e^{i\pi/4} + |\downarrow_{z}\rangle|\uparrow_{z}\rangle 
    e^{-i\pi/4}) \\
    |r_{2}\rangle & = & 
    \frac{1}{\sqrt{2}}|\uparrow_{z}\rangle|\uparrow_{z}\rangle - 
    \frac{1}{2}(|\uparrow_{z}\rangle|\downarrow_{z}\rangle 
    e^{i\pi/4} + |\downarrow_{z}\rangle|\uparrow_{z}\rangle 
    e^{-i\pi/4}) \\
    |r_{3}\rangle & = & 
    \frac{1}{\sqrt{2}}|\downarrow_{z}\rangle|\downarrow_{z}\rangle + 
    \frac{1}{2}(|\uparrow_{z}\rangle|\downarrow_{z}\rangle 
    e^{-i\pi/4} + |\downarrow_{z}\rangle|\uparrow_{z}\rangle 
    e^{i\pi/4}) \\
    |r_{4}\rangle & = & 
    \frac{1}{\sqrt{2}}|\downarrow_{z}\rangle|\downarrow_{z}\rangle - 
    \frac{1}{2}(|\uparrow_{z}\rangle|\downarrow_{z}\rangle 
    e^{-i\pi/4} + |\downarrow_{z}\rangle|\uparrow_{z}\rangle 
    e^{i\pi/4})
\end{eqnarray}

Note that:
\begin{eqnarray}
    |\mbox{pre}\rangle & = & 
    \frac{1}{\sqrt{2}}(|\uparrow_{z}\rangle|\uparrow_{z}\rangle + 
    |\downarrow_{z}\rangle|\downarrow_{z}\rangle \\
                       & = & 
    \frac{1}{\sqrt{2}}(|\uparrow_{x}\rangle|\uparrow_{x}\rangle + 
    |\downarrow_{x}\rangle|\downarrow_{x}\rangle \\
                       & = & 
    \frac{1}{\sqrt{2}}(|\uparrow_{y}\rangle|\downarrow_{y}\rangle + 
    |\downarrow_{y}\rangle|\uparrow_{y}\rangle \\
                       & = & 
    \frac{1}{2}(|r_{1}\rangle + |r_{2}\rangle + |r_{3}\rangle + 
    |r_{4}\rangle)
\end{eqnarray}
    
    Alice can now assign values to the outcomes of Bob's spin measurements  
via the ABL-rule,
whether Bob measured $\sigma_{x}$, $\sigma_{y}$, or $\sigma_{z}$,
based on the post-selections $|r_{1}\rangle$, $|r_{2}\rangle$, 
$|r_{3}\rangle$, or $|r_{4}\rangle$, according to Table~\ref{table}.
\begin{table}[ht]
    \begin{center}
  $
    \begin{array}{r|ccc}
	      & \sigma_{x} & \sigma_{y} & \sigma_{z} \\ \hline
        r_{1} & \uparrow & \uparrow & \uparrow \\ 
        r_{2} & \downarrow & \downarrow & \uparrow \\ 
        r_{3} & \uparrow & \downarrow & \downarrow \\ 
        r_{4} & \downarrow & \uparrow & \downarrow 
     \end{array}
  $
  \end{center}
    \caption{\protect $\sigma_{x}$, \protect $\sigma_{y}$, \protect 
    $\sigma_{z}$ measurement outcomes correlated with eigenvalues of R}
    \label{table}
\end{table}

Consider, now, the following protocol for bit commitment based on the 
Vaidman-Aharonov-Albert case. Alice prepares $n$ copies of the Bell 
state $ 
    |\mbox{pre}\rangle = 
    \frac{1}{\sqrt{2}}(|\uparrow_{z}\rangle_{A}|\uparrow_{z}\rangle_{C} + 
    |\downarrow_{z}\rangle_{A}|\downarrow_{z}\rangle_{C}
$. She keeps the ancillas and sends the channel particles to Bob in 
sequence. Bob measures either $\sigma_{x}$, $\sigma_{y}$, or 
$\sigma_{z}$ chosen randomly on a channel particle, records the 
outcome, and returns the particle to Alice before she sends him the 
next channel particle in the sequence. Alice measures the observable 
$R$ on each channel particle she receives back from Bob. 

The commitment is made as follows: After the sequence of 
measurements, Bob announces the indices in the sequence for which he 
obtained a `$\uparrow$' outcome for his measurements (without announcing 
whether he measured  $\sigma_{x}$, $\sigma_{y}$, or $\sigma_{z}$). The 
remaining elements in the sequence are discarded. 
Alice can now divide the $\uparrow$-sequence into two subsequences of 
approximately equal length (for large $n$): the subsequence $S_{1}$ 
for which she obtained the outcome $r_{1}$ for $R$, and the complementary 
subsequence $S_{234}$ for which she obtained the outcome $r_{2}$, 
$r_{3}$, or $r_{4}$. If Alice commits to 0, she announces the indices 
of the subsequence $S_{234}$ and proves her commitment at the final 
stage, when she reveals her commitment, by her ability to announce 
(from Table~\ref{table}),
for each element in the subsequence, the observable that Bob measured, 
either $\sigma_{x}$, $\sigma_{y}$, or $\sigma_{z}$. If she commits to 
1, she announces the indices of the subsequence $S_{1}$ 
and proves her commitment by her ability to announce, for each element 
in the \textit{complementary} subsequence $S_{234}$, the observable 
that Bob measured.

At first sight, it might appear that this protocol is not of the sort 
covered by the bit commitment theorem. To see that it is, suppose that 
instead of following the protocol and actually choosing one of 
$\sigma_{x}$, $\sigma_{y}$, or $\sigma_{z}$, performing the 
measurement, and obtaining a determinate outcome, Bob implements an 
EPR cheating strategy with a quantum die ancilla with three orthogonal 
states $|d_{x}\rangle$, $|d_{y}\rangle$, $|d_{z}\rangle$ 
corresponding to the choice of spin observable $\sigma_{x}$, $\sigma_{y}$, 
$\sigma_{z}$. Then the state of the composite system consisting of 
Alice's ancilla, the channel particle, and Bob's die and pointer 
ancillas is:
\begin{eqnarray}
    \lefteqn{|\Psi\rangle = 
    \frac{1}{\sqrt{3}}|d_{x}\rangle_{B}(\frac{1}{\sqrt{2}}
    |\uparrow_{x}\rangle_{A}|\uparrow_{x}\rangle_{C}
    |p_{\uparrow}\rangle_{B} + \frac{1}{\sqrt{2}}|\downarrow_{x}\rangle_{A}
    |\downarrow_{x}\rangle_{C}|p_{\downarrow}\rangle_{B})} \nonumber 
    \\
    & & \mbox{} + \frac{1}{\sqrt{3}}|d_{y}\rangle_{B}(\frac{1}{\sqrt{2}}
    |\uparrow_{y}\rangle_{A}
    |\downarrow_{y}\rangle_{C}
    |p_{\downarrow}\rangle_{B} + \frac{1}{\sqrt{2}}|\downarrow_{y}\rangle_{A}
    |\uparrow_{y}\rangle_{C}|p_{\uparrow}\rangle_{B}) \nonumber 
    \\
    & & \mbox{} + \frac{1}{\sqrt{3}}|d_{z}\rangle_{B}(\frac{1}{\sqrt{2}}
    |\uparrow_{z}\rangle_{A}
    |\uparrow_{z}\rangle_{C}
    |p_{\uparrow}\rangle_{B} + \frac{1}{\sqrt{2}}|\downarrow_{z}\rangle_{A}
    |\downarrow_{z}\rangle_{C}|p_{\downarrow}\rangle_{B})  
\end{eqnarray}

To announce `$\uparrow$,' Bob measures the pointer ancilla for $p_{\uparrow}$ or 
$p_{\downarrow}$, which projects 
$|\Psi\rangle$ onto:
\begin{equation}
    |\uparrow\rangle = 
    |p_{\uparrow}\rangle_{B}\frac{1}{\sqrt{3}}(|d_{x}\rangle_{B}
    |\uparrow_{x}\rangle_{A}
    |\uparrow_{x}\rangle_{C} + 
    |d_{y}\rangle_{B}|\downarrow_{y}\rangle_{A}
    |\uparrow_{y}\rangle_{C} +
    |d_{z}\rangle_{B}|\uparrow_{z}\rangle_{A}
    |\uparrow_{z}\rangle_{C})
\end{equation}
or
\begin{equation}
    |\downarrow\rangle = 
    |p_{\downarrow}\rangle_{B}\frac{1}{\sqrt{3}}
    (|d_{x}\rangle_{B}|\downarrow_{x}\rangle_{A}
    |\downarrow_{x}\rangle_{C} + 
    |d_{y}\rangle_{B}|\uparrow_{y}\rangle_{A}
    |\downarrow_{y}\rangle_{C} + 
    |d_{z}\rangle_{B}|\downarrow_{z}\rangle_{A}
    |\downarrow_{z}\rangle_{C})
\end{equation}
with probability $\frac{1}{2}$.
Note that this enables Bob to announce the `$\uparrow$' outcomes 
without actually measuring $\sigma_{x}$, $\sigma_{y}$, or 
$\sigma_{z}$! In effect, he has a quantum computer that computes `$\uparrow$' 
or `$\downarrow$' for the quantum disjunction `$\sigma_{x}$ or $\sigma_{y}$ 
or 
$\sigma_{z}$.'  

The state $|\uparrow\rangle$ can be expressed in terms of 
$R$-eigenstates:
\begin{eqnarray}
    \lefteqn{|\uparrow\rangle = \frac{1}{\sqrt{3}}(\frac{1}{\sqrt{2}}
    |r_{1}\rangle_{A} + \frac{1}{\sqrt{2}}
    |r_{3}\rangle_{A})|d_{x}\rangle_{B}|p_{\uparrow}\rangle_{B}} 
    \nonumber \\
    & & \mbox{} + \frac{1}{\sqrt{3}}(\frac{1}{\sqrt{2}}|r_{1}\rangle_{A} + 
    \frac{1}{\sqrt{2}}
    |r_{4}\rangle_{A})|d_{y}\rangle_{B}|p_{\uparrow}\rangle_{B}
    \nonumber \\
    & & \mbox{} + \frac{1}{\sqrt{3}}(\frac{1}{\sqrt{2}}|r_{1}\rangle_{A} + 
    \frac{1}{\sqrt{2}}|r_{2}\rangle_{A})|d_{z}\rangle_{B}|p_{\uparrow}\rangle_{B}
\end{eqnarray}
and rewritten as:
\begin{eqnarray}
    \lefteqn{|\uparrow\rangle = \frac{1}{\sqrt{2}}
    |r_{1}\rangle_{A}(\frac{1}{\sqrt{3}}|d_{x}\rangle_{B} + 
    \frac{1}{\sqrt{3}}|d_{y}\rangle_{B} + \frac{1}{\sqrt{3}}
    |d_{z}\rangle_{B}) |p_{\uparrow}\rangle_{B}}
    \nonumber \\
    & & \mbox{} + \frac{1}{\sqrt{2}}(\frac{1}{\sqrt{3}}
    |r_{3}\rangle_{A}|d_{x}\rangle_{B} 
    + \frac{1}{\sqrt{3}}|r_{4}\rangle_{A}|d_{y}\rangle_{B} + 
    \frac{1}{\sqrt{3}}|r_{2}\rangle_{A}|d_{z}\rangle_{B})
    |p_{\uparrow}\rangle
\end{eqnarray}

Evidently, after Alice measures the observable $R$ on the channel 
particles in the `$\uparrow$' subsequence and announces either the 
subsequence $S_{234}$ for which she obtained the eigenvalues 
$r_{2}$, $r_{3}$, or $r_{4}$
corresponding to the 0-commitment, or the subsequence $S_{1}$ for which she 
obtained the eigenvalue $r_{1}$
corresponding to the 1-commitment, Bob's density operator for 
the channel particles (obtained by tracing over Alice's ancillas and 
Bob's ancillas) will be either: 
\begin{equation}
    W_{B}(0) = \frac{1}{3}(|d_{x}\rangle_{B}\mbox{}_{B}\langle d_{x}| +
    |d_{y}\rangle_{B}\mbox{}_{B}\langle d_{y}| + 
    |d_{z}\rangle_{B}\mbox{}_{B}\langle d_{z}|)
\end{equation}
for the subsequence $S_{234}$, or:
\begin{equation}
    W_{B}(1) = \frac{1}{3}
    (|d_{x}\rangle_{B} + |d_{y}\rangle_{B} + |d_{z}\rangle_{B})
    (\mbox{}_{B}\langle d_{x}| +\mbox{}_{B}\langle d_{y}| + 
    \mbox{}_{B}\langle d_{z}|)
\end{equation}
for the subsequence $S_{1}$. (More precisely, these are the density 
operators for a single channel particle. The density operator for the 
sequence of channel particles is in each case a tensor product of 
the relevant operator 
over the elements of the sequence.) 
But these density operators are 
distinguishable: $W_{0}$ is the density operator of an equal weight mixture of 
pure states $|d_{x}\rangle_{B}$, $|d_{y}\rangle_{B}$, and 
$|d_{z}\rangle_{B}$,
while  $W_{1}$ is the density operator of the pure state
$
\frac{1}{\sqrt{3}}|d_{x}\rangle_{B} + 
    \frac{1}{\sqrt{3}}|d_{y}\rangle_{B} + \frac{1}{\sqrt{3}}
    |d_{z}\rangle_{B}
$. So Bob can cheat---the protocol is insecure. 

Now, suppose we assume that Bob is forced to make a determinate 
choice of which spin component observable to measure for each channel 
particle, and actually 
perform the measurements and record the outcomes. Then it is clear 
that both subsequences $S_{1}$ and $S_{234}$ 
will be characterized by the same equal weight mixture of 
pure states $|d_{x}\rangle_{B}$, $|d_{y}\rangle_{B}$, and 
$|d_{z}\rangle_{B}$. So Bob cannot cheat. But Alice cannot cheat either. 
Of course, by the bit commitment theorem, 
since Alice is in possession of all the channel particles 
at the final stage when she is required to reveal her commitment, 
there exists a unitary transformation in Alice's Hilbert space (which 
now includes the channel particles) that will 
transform the states of the ancilla-channel pairs to $R$-eigenstates 
that conform to Bob's measurement outcomes. But this unitary 
transformation depends on the outcomes of Bob's measurements, which 
are unknown to Alice. Essentially, Alice would have to transform the 
state $|r_{1}\rangle$ for each element in the declared subsequence or 
the 
complementary subsequence to the state 
$|r_{2}\rangle$, $|r_{3}\rangle$, or $|r_{4}\rangle$, 
corresponding to Bob's measurement outcome 
for that element,
in order to successfully change her commitment without Bob being able 
to detect her cheating. There exists a unitary transformation that 
Alice can implement to achieve this result, but she cannot know what 
unitary transformation to employ. So the protocol is secure, subject to the 
assumption that Bob cannot apply an EPR cheating strategy. 

The question raised at the beginning of this section can now be put 
more concretely. In the above protocol, if Bob is honest and does not 
apply an EPR strategy, then neither party can cheat. If he 
applies the strategy, then he gains the advantage. Can there be a bit 
commitment protocol that is 
similar to the above protocol, except that the application of an EPR 
strategy by Bob at a certain stage of the protocol would give 
Alice the advantage, rather than Bob, while conforming to the protocol 
would ensure that neither party could cheat? If there were such a 
protocol, then Bob would, in effect, be forced to conform to the 
protocol and avoid the EPR strategy, and unconditionally 
secure bit commitment would be possible. 

In fact, the impossibility of such a protocol follows from the theorem 
itself. 
Suppose there were such a protocol. That is, suppose that 
if Bob applies an EPR strategy then $W_{B}(0) = W_{B}(1)$, so 
by the theorem there exists a unitary transformation $U$ in Alice's 
Hilbert space that will transform $|0\rangle$ to $|1\rangle$. Alice 
must know this $U$ because it is uniquely determined by Bob's 
deviation from the protocol according to an EPR strategy 
that keeps all disjunctions at the quantum level as linear 
superpositions. Suppose also that if, instead, Bob is honest and follows the 
protocol (so that there is a determinate choice for every disjunction 
over possible operations or possible measurement outcomes), then 
$W_{B}(0) = W_{B}(1)$,
but the unitary transformation in Alice's Hilbert space that allows 
her to transform $|0\rangle$ to $|1\rangle$ depends on Bob's choices 
or measurement outcomes, which are unknown to Alice. 

Now the crucial point to note is that the information available in Alice's 
Hilbert space must be the same whether Bob follows the protocol and 
makes determinate choices and obtains determinate measurement 
outcomes before Alice applies the unitary transformation $U$ 
that transforms $|0\rangle$ to $|1\rangle$, or whether he 
deviates from the protocol via an EPR strategy 
in which he implements corresponding entanglements with his ancillas 
to keep choices and measurement outcomes at the quantum level before 
Alice applies the transformation $U$, and only makes these choices and 
measurement outcomes determinate at the final stage of the protocol by 
measuring his ancillas. There 
can be no difference for Alice because Bob's measurements on his 
ancillas and any measurements or operations 
that Alice might perform take place in 
different Hilbert spaces, so the operations commute. If Alice's 
density operator (obtained by tracing over Bob's ancillas), which 
characterizes the statistics of measurements that Alice can perform in 
her part of the universe, were different depending on whether or not 
Bob actually carried out the required measurements, as opposed to 
keeping the alternatives at the quantum level by implementing 
corresponding 
entanglements with ancillas, then it would be possible to use this 
difference to signal superluminally. Actual measurements by Bob on his 
ancillas 
that selected alternatives in the entanglements 
as determinate would instantaneously alter the information available 
in Alice's part of the universe. 

It follows that in the hypothetical bit commitment protocol we are considering, 
the unitary transformation $U$ in 
Alice's Hilbert space that transforms $|0\rangle$ to $|1\rangle$ 
must be the same transformation in the honest scenario as in the 
cheating scenario. But we 
are assuming that the transformation in the honest scenario is unknown to Alice 
and depends on Bob's measurement outcomes, while the transformation in 
the cheating scenario is unique and known to Alice. So there can be no such 
protocol: the deviation from the protocol by an EPR strategy 
can never place Bob in 
a worse position than following the protocol honestly.  

The argument can be put formally in terms of the theorem as follows: 
The cheating scenario produces one of two alternative pure 
states $|0\rangle_{c}$ or $|1\rangle_{c}$ in 
$\mathcal{H}_{A}\otimes\mathcal{H}_{B}$ (`$c$' for `cheating strategy). 
Since the reduced density 
operators in $\mathcal{H}_{B}$:
\begin{eqnarray}
    W^{(c)}_{B}(0) & = & Tr_{A}|0\rangle_{c}\mbox{}_{c}\langle 0|
    \nonumber \\
    W^{(c)}_{B}(1) & = & Tr_{A}|1\rangle_{c}\mbox{}_{c}\langle 1|
\end{eqnarray}
are required by assumption to be the same:
\begin{equation}
    W^{(c)}_{B}(0) = W^{(c)}_{B}(1)
\end{equation}
the states $|0\rangle_{c}$ and $|1\rangle_{c}$ can be expressed in
biorthogonal decomposition as:
\begin{eqnarray}
    |0\rangle_{c} & = & \sum_{i}\sqrt{c_{i}}|a_{i}\rangle |b_{i}\rangle
    \nonumber \\
    |1\rangle_{c} & = & \sum_{i}\sqrt{c_{i}}|a'_{i}\rangle |b_{i}\rangle
\end{eqnarray}
where the reduced density operators in $\mathcal{H}_{A}$:
\begin{eqnarray}
    W^{(c)}_{A}(0) = Tr_{B}|0\rangle_{c}\mbox{}_{c}\langle 0|
    & = & \sum_{i}|c_{i}|a_{i}\rangle\langle a_{i}|
    \nonumber \\ 
    W^{(c)}_{A}(1) = Tr_{B}|1\rangle_{c}\mbox{}_{c}\langle 1|
    & = & \sum_{i}|c_{i}|a'_{i}\rangle\langle a'_{i}|
\end{eqnarray}
are different:
\begin{equation}
    W^{(c)}_{A}(0) \neq W^{(c)}_{A}(1)
\end{equation} 

It follows that there exists a unitary 
operator $U_{c} \in \mathcal{H}_{A}$ defined by the spectral 
representations of $W_{A}^{(c)}(0)$ and $W_{A}^{(c)}(1)$: 
\begin{equation}
    \{|a_{i}\rangle\} \stackrel{U_{c}}{\longrightarrow} \{|a'_{i}\rangle\}
\end{equation}
such that:
\begin{equation}
    |0\rangle_{c} \stackrel{U_{c}}{\longrightarrow} 
    |1\rangle_{c}
\end{equation}

The honest scenario produces one of two alternative pure states 
$|0\rangle_{h}$ and $|1\rangle_{h}$ in 
$\mathcal{H}_{A}\otimes\mathcal{H}_{B}$ (`$h$' for `honest scenario'), 
where the pair 
$\{|0\rangle_{h}, |1\rangle_{h}\}$ depends on Bob's choices and 
the outcomes of his measurements. 

By assumption, as in the cheating scenario, the reduced density operators 
$W_{B}^{(h)}(0)$ and $W_{B}^{(h)}(1)$ in $\mathcal{H}_{B}$ are the 
same:
\begin{equation}
    W^{(h)}_{B}(0) = W^{(h)}_{B}(1)
\end{equation}
which entails the existence of a 
unitary operator $U_{h} \in \mathcal{H}_{A}$ such that:
\begin{equation}
    |0\rangle_{h} \stackrel{U_{h}}{\longrightarrow} 
    |1\rangle_{h}
\end{equation}
where $U_{h}$ depends on Bob's choices and measurement outcomes. 

Now, the difference between the honest scenario and the cheating scenario is 
undetectable in $\mathcal{H}_{A}$, which means that 
the reduced density operators 
in $\mathcal{H}_{A}$ are the same in the honest scenario as in the 
cheating scenario:
\begin{eqnarray}
     W^{(h)}_{A}(0) & = & W^{(c)}_{A}(0)
     \nonumber \\
     W^{(h)}_{A}(1) & = & W^{(c)}_{A}(1)
\end{eqnarray}
Since $U_{h}$ is defined by the spectral representations of 
$W^{(h)}_{A}(0)$ and $W^{(h)}_{A}(1)$, it follows that 
$U_{h} = U_{c}$. But we are assuming that 
$U_{h}$ depends on Bob's 
choices and measurement outcomes, while $U_{c}$ is uniquely defined 
by Bob's EPR strategy, in which there are no determinate 
choices or measurement outcomes. Conclusion: there can be no bit 
commitment protocol in which neither Alice nor Bob can cheat if Bob 
honestly follows the protocol, but Alice can cheat
if Bob deviates from the protocol via an EPR strategy. 
If neither Bob nor Alice can cheat in the honest scenario, then Bob and not 
Alice  must 
be able to cheat in the cheating scenario.

A similar argument rules out a protocol in which neither party 
can cheat if Bob is honest (as above), but if Bob follows an EPR 
strategy, then $W_{B}(0) \approx W_{B}(1)$, so Bob has some probability of 
cheating successfully, but Alice has a 
greater probability of cheating successfully than Bob. Again, the 
unitary transformation $U_{c}$ that would allow Alice to cheat with a certain 
probability of success if Bob 
followed an EPR strategy would also have to allow Alice to 
cheat successfully if Bob were honest. But the supposition is that 
Alice cannot cheat if Bob is honest, because the unitary 
transformation $U_{h}$ in that case depends on Bob's choices and 
measurement outcomes, which are unknown to Alice. It follows that 
there can be no such protocol. 

So there is no loophole -- not even in the extended sense: 
following an EPR cheating strategy can never be 
disadvantageous to the cheater. 
Unconditionally secure quantum bit commitment (in the sense of the 
theorem) really 
is impossible.

\section*{Acknowledgements}
This work was partially supported by a University of Maryland 
General Research Board leave fellowship. My understanding of the 
quantum bit commitment problem owes 
a very large debt to discussions and email correspondence with 
Adrian Kent. 
Thanks also to Gilles Brassard, Dominic Mayers, Rob Clifton, and David MacCallum
for illuminating discussions.

\end{document}